\documentclass[11pt]{article}
\usepackage{amssymb,amsmath,bm,a4}

\newcommand{\eeq}{\end{equation}}\newcommand{\beq}{\begin{equation}}
\newcommand{\ba}{\begin{array}}
\newcommand{\ea}{\end{array}}
\newcommand{\bea}{\begin{eqnarray}}
\newcommand{\eea}{\end{eqnarray}}

\newcommand{\lae}{\lambda_e}
\newcommand{\lap}{\lambda_p}
\newcommand{\lan}{\lambda_n}

\begin{document}


\title{ 
Composition-dependent long range forces from varying $m_p/m_e$}

\author{Thomas Dent\thanks{email:\,dent@thphys.uni-heidelberg.de},\\
	{\em Theoretische Physik, 16 Philosophenweg, Heidelberg 69120 GERMANY}
	}


\date{\today}

\maketitle

\begin{abstract}
\noindent We relate possible cosmological variations in the mass ratio 
$\mu\equiv m_p/m_e$ and the fine structure constant $\alpha$ to 
long-range composition-dependent forces mediated by a scalar field. 
The differential acceleration $\eta$ in E{\" o}tv{\" o}s-type experiments 
is bounded below by $10^{-14}$, except in cases where one or more scalar 
couplings vanish. We consider what values for these couplings could 
arise from unified theories. By considering the contribution of the 
scalar field to the cosmological energy density we use bounds on $\eta$ 
to put upper bounds on the current rate of change of $\mu$ and $\alpha$. 
\end{abstract}



\section{Introduction}\noindent
The constancy of the coupling strengths and particle masses in the Standard 
Model Lagrangian is an assumption that should be tested. If a variation 
existed it would violate the principle of Local Position Invariance contained
in the Einstein equivalence principle. Nevertheless consistent, 
relativistically covariant theories with varying couplings can be written 
down by including scalar degrees of freedom, in which the variation arises 
due to the cosmological solution for the scalar(s). See for example 
\cite{Bekenstein,OliveP,Wetterich02,BarrowElectron}.

There are also many situations in cosmology where gravitational effects appear
to behave differenty from what a naive application of GR would produce, which
may also motivate the existence of cosmological scalar fields. The first
is inflation (see for example \cite{WMAPcosmo}), which seems to require a 
scalar field with a peculiar kind of potential. Closely related, though 
happening on an immensely slower timescale, is the current apparent 
accelerated expansion of the Universe \cite{supernovae,WMAPcosmo}, which 
may also be related to scalar field dynamics \cite{cosmonquint,Wetterich02}.

The violation of the Einstein equivalence principle makes itself via scalar 
mediated forces, thus gravity becomes a generalised scalar-tensor theory. 
So far, precision tests of gravity have revealed no deviation from General 
Relativity \cite{Will01,Nordtvedt02}, in particular concerning the question 
we will discuss here,
the equivalence of gravitational and inertial mass (universality of free
fall) \cite{Su:1994gu}.
The motivation for continuing to consider deviations from GR then comes 
into question. 

The current motivation comes from astrophysical measurements of various 
coupling strengths and mass ratios deduced from atomic and molecular
absorption (or emission) wavelengths. These probe back to redshifts up to 
about 4 and have now reached the sensitivity of $10^{-5}$ or better; 
moreover some measurements of the fine structure constant $\alpha$ 
\cite{Webbalpha}, and most recently the proton-electron mass ratio $\mu
\equiv m_p/m_e$ \cite{mu} show potentially significant deviations. The fine 
structure constant has been the subject of debate, with other recent studies 
obtaining negative results \cite{novary,levshakov}, some at a precision 
greater than or equal to previous evidence for variation. 

Bounds on variation of $\alpha$ from nuclear physics effects 
\cite{Oklo,Olive_et_al} have also been derived; however they are subject
to considerable uncertainty if one also allows other quantities such as 
quark masses to vary (see for example \cite{FlambaumS}).

It was pointed out in \cite{DvaliZ} that a fractional variation of 
$\alpha$ on the order of $10^{-5}$ required a scalar coupling to 
electromagnetic energy large enough such that the differential acceleration
of two bodies in the gravitational field of the Sun (including scalar 
forces) is of order $10^{-18}$, which may be detectable with the STEP 
experiment. A more detailed analysis of scalar field models in \cite{OliveP}, 
including also the electrostatic energy of nuclei, found a rather 
more restrictive bound from non-universal free fall, generally referred 
to as Weak Equivalence Principle (WEP) violation, although. Other 
works have related a variation in $\alpha$ to scalar field models which 
describe the current acceleration of the Universe 
\cite{Anchordoqui:2003ij,Parkinson:2003kf,Doran:2004ek}.


As pointed out in \cite{Wetterich02}, one reason for believing the
result of \cite{DvaliZ} to be an underestimate of WEP violation is that
the authors did not include possible scalar couplings to other Standard 
Model sectors which are likely in unified models (see also 
\cite{C+O,Calmet,Langacker,us,Dine,me03}). Clearly also a scalar 
coupling to electromagnetism alone cannot produce a variation in $\mu$ 
at the same level as that in $\alpha$.

Hence in this paper we redo and extend the analysis to include other 
possible types of scalar coupling through which varying $\alpha$ and 
$\mu$ can be related to WEP violation. Current limits on differential 
gravitational acceleration are at the level of $10^{-12}$ -- $10^{-13}$ 
\cite{Su:1994gu} 
for the parameter $\eta$ defined as
\beq \label{etadef}
 \eta \equiv 2\frac{|a_1-a_2|}{|a_1+a_2|} 
\eeq
where two test bodies of different composition have accelerations $a_1$, 
$a_2$ towards a known gravitational source.

In the first section of the paper we make a preliminary estimate of the 
size of differential acceleration $\eta$ due to $\phi$-mediated forces,
allowing for varying $\mu$, and find it may be orders of magnitude greater 
than $10^{-18}$, and possibly visible with the MICROSCOPE experiment
with sensitivity $10^{-15}$.
Exceptions occur if some scalar couplings vanish exactly.

In the second section we consider what scalar couplings consistent with 
varying $\mu$ may emerge from unified models both with and without 
supersymmetry, using the detailed analysis of \cite{me03}. We find 
it unlikely that the scalar coupling to any particle will vanish. In 
fact some unification scenarios imply that WEP violation may be at 
the current limits of detection, or even be ruled out by E{\" o}tv{\" o}s-type 
experiments. 
Only a restricted range of unified scenarios can be consistent with 
reported variations of $\alpha$ and $\mu$, if these are to arise 
from a scalar field varying over cosmological time.

In the last section we examine the cosmological evolution of the scalar 
field in more detail
and suggest a direct test of unified scenarios by comparing WEP violation 
with atomic clock measurements of the constancy of fundamental parameters.

\subsection{Consistency of ``varying constant'' observations}

Clearly, the experimental situation on possible variation of $\alpha$ is
currently unsatisfactory, while that for $\mu$ requires further 
independent testing. We discuss two points: first, whether nonzero 
variations in both $\alpha$ and $\mu$ are inconsistent with other 
measurements of dimensionless fundamental quantities; second, whether 
recent results on $\alpha$ are directly comparable.

The quantities $y\equiv \alpha^2 g_p$ and $x \equiv \alpha^2 g_p m_e/m_p$
have been measured with good precision using quasar absorption 
\cite{Murphy01,Tzanavaris}. From two systems at redshifts below $0.7$ 
the fractional variation of $y$ is bounded at the level of $5\times 
10^{-6}$, hence is not inconsistent with the claim $\Delta \alpha = 
(-0.57 \pm 0.11) \times 10^{-5}$ (made for systems over a wide range in
redshift) given constant $g_p$.~\footnote{Possible variation of $g$-factors
was discussed in \cite{Dmitriev:2002kv}; any effect on $g_p$ is expected 
to be small.} The result for $x$ indicates a slight but
not significant {\em increase} into the past, $\Delta x/x = (1.2 \pm 1.0)
\times 10^{-5}$. Again given constant $g_p$, this is difficult to 
reconcile with a decrease in $\alpha$ and an increase in $\mu$ as 
claimed in \cite{mu} at the level $\Delta \mu/\mu = (2.4 \pm 0.6) \times
10^{-5}$. Hence we cannot assume that both $\alpha$ and $\mu$ have
exactly the claimed nonzero variation. Rather we will make estimates 
assuming that at least one nonzero variation exists (usually $\mu$)
at the claimed level.

The most recent negative results for $\Delta\alpha$ are based on very few 
absorption systems, the improved sensitivity being ascribed to more accurate 
spectroscopy. The Webb group results by contrast involve a sample of about 
150 systems.\footnote{Analysis of more recent observations is ongoing 
(M.~Murphy, private communication).} This makes it possible to compare the 
estimated error in each system
statistical scatter over the sample, and carry out checks against systematic 
error. One outstanding question is the statistical error due to fitting the 
velocity profile of the absorbing system; see for example the first reference 
of \cite{levshakov}. 
It may require a large number of systems to be analyzed to settle the question.

\section{Preliminary estimate of $\eta$}
The estimate proceeds along the same lines as \cite{DvaliZ}, but with some 
important differences. In particular the electron content of matter must be 
explicitly included since the variation of $\mu$ involves $m_e$. 

We posit a light neutral scalar field $\phi$ which is canonically normalised,  
redefining the origin such that the value of $\phi$ vanishes today, and work
in the Einstein frame. It will be convenient to use a dimensionless scalar
$\bar{\phi}\equiv \phi/\bar{M}_P$ where $\bar{M}_P$ is the reduced Planck 
mass $(8\pi G)^{-1/2}$. The scalar couples to matter in the low momentum 
limit via effective operators 
\beq \label{phimatter}
 \mathcal{L}_{\phi m} = - m_{e,0} \left[1 + \lae\bar{\phi} \right] \bar{e}e 
 - m_{p,0} \left[1 + \lap\bar{\phi} \right] \bar{p}p
 - m_{n,0} \left[1 + \lan\bar{\phi} \right] \bar{n}n
\eeq
where $m_{e,0}$ {\em etc.}, are the present-day values of particle masses,
and to electromagnetism via
\beq \label{phiem}
 \mathcal{L}_{\phi em} = -\frac{1}{4} \left[1+\lambda_{em} \bar{\phi} \right]^{-1}
 F_{\mu\nu} F^{\mu\nu}
\eeq
thus the fine structure constant is given by $\alpha = \alpha_0 
(1+\lambda_{em} \bar{\phi})$.
Note that $\lap$ and $\lan$ can differ from one another only 
due to light quark masses and electromagnetic effects, {\em i.e.}\ isospin 
violation. We may reexpress the nucleon mass terms as
\beq \label{nucmass}
 m_{N} \left[1 + \lambda_i \bar\phi \right] (\bar{p}p +\bar{n}n)
 + \frac{m_{N}}{2} \left[\frac{\delta m_{np}}{m_{N}} 
 + \lambda_v \bar{\phi} \right] (-\bar{p}p +\bar{n}n)
\eeq
where $m_N$ is an averaged nucleon mass (we drop the ``0'' suffix except 
in case of ambiguity), $\delta m_{np}$ is the nucleon mass difference and 
\[ 
 \lambda_i \equiv \frac{\lap m_p + \lan m_n}{m_N},\ \lambda_v \equiv 
 \frac{-\lap m_p +\lan m_n}{m_N}.
\]
Note that this form gives only the leading linear dependence on 
$\bar{\phi}$. We could be more general and replace $\lambda_e$ and other
couplings with general functions $\lambda_e(\bar{\phi})$ (etc.) defined via
\[
 \frac{\partial \ln (m_e/M_P)}{\partial \bar{\phi}} \equiv 
 \lambda_e(\bar{\phi})
\]
allowing a nonlinear dependence, which may be likely if the value of 
$\bar{\phi}$ changes by more than a small fraction.

For a given test object we have $n_e = n_p$ electrons and protons and $n_n$ 
neutrons, with the ``proton fraction'' $f_p$ defined as \mbox{$n_p/(n_p+n_n)$},
equal to the mass averaged value of $Z/A$, and ``neutron fraction'' 
$f_n\equiv 1-f_p$. The nuclear electromagnetic
binding energy also makes a contribution \mbox{$f_{em} m_N Z(Z-1) A^{-1/3}$} to 
the mass, where $f_{em}\simeq 7\times 10^{-4}$.
The ratio of the $\phi$-mediated acceleration of a test body $a_\phi$ to the 
gravitational acceleration $a_{gr}$ is then
\begin{multline} \label{aphi}
 \frac{a_\phi}{a_{gr}} = 
 \left( \frac{ f_p^s (\lae m_e + \lap m_p) + f_n^s \lan m_n } {m_N} + 
 f_{em} \lambda_{em} \frac{\langle Z(Z-1) A^{-1/3} \rangle^s}{\langle A\rangle^s} \right)
 \times \\
 \left( \frac{ f_p (\lae m_e + \lap m_p) + f_n \lan m_n } {m_N} + 
 f_{em} \lambda_{em} \frac{\langle Z(Z-1) A^{-1/3} \rangle}{\langle A\rangle} \right) 
\end{multline}
where the superscript $s$ 
refers to the common source. If, as in some experiments, this is the Sun then
the electromagnetic contribution to the source coupling is negligible. 
Note that if electrons are in relativistic motion their scalar charge is 
``diluted'' by a boost factor, as derived for example in 
\cite{BarrowElectron}, but to obtain conservative bounds on scalar 
couplings we will take the nonrelativistic limit.

For two test bodies possessing different $f_p$, $Z$ and $A$ one can 
determine the ratio
\beq 
 \eta 
 \simeq \frac{|a_{\phi,1}-a_{\phi,2}|}{a_{gr}}
\eeq
where 
$a_\phi \ll a_{gr}$. 
We then obtain
\begin{multline} \label{etaitolambda}
 \eta \simeq \left( f_p^s \left[ \lae \frac{m_e}{m_N} + \lap \right] + f_n^s \lan 
 + \lambda_{em} f_{em} \frac{\langle Z(Z-1) A^{-1/3}\rangle^s}{\langle A\rangle^s} \right)
 \times \\
 \left( \left[ \lae \frac{m_e}{m_N} + \lap \frac{m_p}{m_N} - \lan \frac{m_n}{m_N} \right] 
 \Delta_{12} f_p 
 + \lambda_{em} f_{em} \Delta_{12} \frac{\langle Z^2 A^{-1/3}\rangle}{\langle A\rangle} \right)
 \\
 \simeq \left(\frac{\lae}{1837} + \frac{\lambda_i}{2} - \frac{\lambda_v}{2} \right) 
 \left(  \left[ \frac{\lae}{1837} -\lambda_v \right] \Delta_{12} f_p
 + \lambda_{em} f_{em} \Delta_{12} \frac{\langle Z^2 A^{-1/3} \rangle}{\langle A\rangle} \right)
\end{multline}
where the notation $\Delta_{12}$ denotes the difference between the 
two test bodies and we take them to have $Z\gg 1$.
The last line of Eq.~(\ref{etaitolambda}) follows since we find the 
strongest bound from an experiment using ``Earth-like'' (mostly iron) 
and ``Moon-like'' (mostly silica) test bodies accelerating towards 
the Sun, for which $f_p^s\simeq 1$; the test bodies differ by $\Delta_{12} 
f_p =0.036$ and $\Delta_{12} ( \langle Z^2 A^{-1/3} \rangle/ 
\langle A\rangle) \simeq -1.0$. 

There are different possibilities for an evolving value of $\phi$ to produce 
a variation in $\mu$. If the variation is entirely due to the 
isospin-conserving coupling $\lambda_i$ then the differential acceleration 
vanishes, but in general one expects all couplings to be nonzero. (Variation 
of $\alpha$ also implies that $\lambda_v$ and $\lambda_{em}$ are nonvanishing.) 
We require $(\lambda_p-\lambda_e)\Delta\bar{\phi}$ to be of order $10^{-5}$, 
where the change $\Delta\phi$ occurs between redshifts $2$ to $3$ and the 
present. If we have a nonlinear dependence of particle masses and $\alpha$
on $\phi$ then the functions $\lambda_{e}(\bar{\phi})$ {\em etc.} must be 
integrated back to the value $\bar{\phi}_1$ at redshift $2$ to $3$ and we 
obtain instead averaged values
\[
 \bar{\lambda}_e = \frac{1}{\Delta \bar{\phi}} 
 \int_{\bar{\phi}_0}^{\bar{\phi}_1} d\bar{\phi}\, \lae(\bar{\phi})
\]
(where we earlier fixed ${\bar{\phi}}_0=0$).
In terms of the isospin conserving and violating $\phi$ couplings we have
\beq \label{muitolambda}
 \Delta \ln \mu = \left(\lambda_i - \frac{1}{2} \lambda_v - \lae\right) \Delta\bar{\phi} 
\eeq
If $\phi$ is not evolving fast, $\Delta \bar{\phi} \lesssim 1$ then at 
least one of these coupling constants should be of magnitude $10^{-5}$ or
greater.

If as in \cite{BarrowElectron} the variation of $\mu$ is entirely due to 
the electron coupling then $\eta$ will be at least $10^{-18}$.
However, if both $\lambda_i$ and $\lae$ are of order $10^{-5}$ then the
lower bound on $\eta$ becomes much larger, a few times $10^{-15}$. If 
$\lambda_i$ happens to vanish, 
then the value of $\eta$ will depend on 
the relation between $\lambda_v$, $\lae$ and $\lambda_{em}$. The 
contribution of electromagnetic self-energy to the proton mass (which 
enters via $\lambda_v$) is tiny and cannot lead to observable changes 
in $\mu$, therefore $\lambda_{em}$ is bounded only by measurements 
of $\alpha$. If we take a variation $\Delta\alpha / \alpha = 0.6\times 
10^{-5}$ then the contribution of $\lambda_{em}$ to $\eta$ is about 
an order of magnitude larger than that of $\lae$, comparing 
$0.036 \lae/1837$ with $7\times 10^{-4}\lambda_{em}$, where $\lae$ is
chosen to give $\Delta\mu/\mu$ of a few times $10^{-5}$.

\section{Scalar couplings in unified models}
In the previous discussion we ignored the underlying physics which should lead 
to relations between the scalar coupling strengths $\lambda_i$, $\lambda_v$, 
$\lae$ and $\lambda_{em}$. In particular the isospin-violating coupling $\lambda_v$ can only 
arise from the coupling of $\phi$ to up and down quark masses or electromagnetic 
self-energy. Now recall that the ratio $m_q/M_P$, where $M_P$ is the Planck mass, 
is mainly due to the hierarchically small value of the Higgs v.e.v.\ $v/M_P$, 
which is equally true of the electron mass: hence even without invoking 
unification it seems very likely that the functional dependence of $m_{u,d}/M_P$ 
on $\phi$ will mirror that of $m_e/M_P$. Symbolically
\beq
 \partial_{\phi} \ln \frac{m_q}{M_P} \simeq \partial_{\phi} \ln \frac{m_e}{M_P}\
 \Rightarrow \ \lambda_{u,d} \simeq \lambda_e.
\eeq
Now the contribution of $m_d-m_u$ to $\delta m_{np}$ is a few MeV, hence the 
value of $\lambda_v$ is suppressed by the ratio $(m_d-m_u)/m_N$, relative 
to the quark couplings, and its contribution to $\eta$ will likely be 
comparable to that of $\lambda_e$.

To be more systematic, we note the following relations valid for small 
variations in particle masses:
\begin{multline}
 \Delta \ln \frac{m_N}{M_P} = \lambda_i \Delta\bar{\phi},\ \ 
 \Delta \ln \frac{\delta m_{np}}{M_P} = \frac{m_N}{\delta m_{np}} \lambda_v \Delta\bar{\phi},\ \
 \Delta \ln \frac{m_e}{M_P} = \lambda_e  \Delta\bar{\phi}
\end{multline}
which will allow us to connect these low-energy quantities to underlying
couplings of the cosmon. We choose to look at models with unification of
gauge couplings in which the cosmon may couple to four quantities: the 
unification mass $M_X$, the unified fine structure constant $\alpha_X$ 
defined at the scale $M_X$, the Higgs v.e.v.\ $v$, and (for supersymmetric 
theories) the soft supersymmetry-breaking masses $\tilde{m}$, which enter 
as thresholds in the RG evolution equations.
Their variations are defined as 
\begin{multline}
 \Delta \ln\frac{M_X}{M_P} = \lambda_M \Delta\bar{\phi},\qquad 
 \Delta \ln\alpha_X = \lambda_X \Delta\bar{\phi}, \\
 \Delta \ln\frac{v}{M_X} = \lambda_H \Delta\bar{\phi},\qquad
 \Delta \ln\frac{\tilde{m}}{M_X} = \lambda_S \Delta\bar{\phi} 
\end{multline}
and in nonsupersymmetric theories we set $\lambda_S=0$. The results of 
\cite{me03} can now be used to relate the nucleon and electron masses
to the underlying quantities. The electron mass varies as 
\beq
\Delta \ln\frac{m_e}{M_P} = {\lambda_M + \lambda_H} \Delta\bar{\phi}
\eeq
thus simply $\lambda_e=\lambda_M+\lambda_H$; the same is true for the 
quark masses (neglecting the variation of Yukawa coupling constants).
As an intermediate result we require the variation of the QCD scale 
$\Lambda_c$ which satisfies
\beq
 \Delta \ln\frac{\Lambda_c}{M_X} \equiv \lambda_\Lambda \Delta\bar{\phi} 
 = \left( \frac{2\pi}{9\alpha_X}\lambda_X + \frac{2}{9} \lambda_H 
  + \frac{4}{9} \lambda_S \right) \Delta\bar{\phi}
\eeq
and the variation of $\alpha$ given by 
\beq
 \Delta \ln\alpha \equiv \lambda_{em}\Delta\bar{\phi} 
  \simeq \left\{ \frac{8\alpha}{3\alpha_X} \lambda_X  
  + \frac{\alpha}{2\pi} \left(\frac{8}{3}\lambda_\Lambda + \lambda_H + 
  \frac{25}{3} \lambda_S  \right) \right\} \Delta\bar{\phi}.
\eeq
Then the averaged nucleon mass coupling is
\beq
 \lambda_i = \lambda_M + \frac{0.54}{\alpha_X}\lambda_X + (0.39 \pm 0.12)\lambda_H + 0.35 \lambda_S
\eeq
where light quark mass contributions, heavy quark and superpartner 
thresholds are accounted for; the uncertainty in the $\lambda_H$ term 
arises from the strangeness content (see for example \cite{Gregory:2005ms}). 
The isospin violating coupling is found from the expression 
\beq
 \delta m_{np} = \left[ 2.05\left( 1+\Delta\ln\frac{m_d-m_u}{M_P} \right) 
  -0.76\left( 1+\Delta\ln\alpha + \Delta\ln\frac{\Lambda_c}{M_P} \right) \right]\, \mbox{MeV}
\eeq
with coefficients derived from \cite{GasserL}; this leads to 
\beq
 \lambda_v \simeq 1.4\times 10^{-3} \left[ \lambda_M - 10.2 \lambda_X 
  + 1.46 \lambda_H - 0.27 \lambda_S \right]
\eeq
where we take $\alpha_X^{-1}\simeq 24$. Note that that the contribution of 
varying $\alpha$ to this mass splitting is always very small compared to that 
of $\Lambda_c$ which is much more sensitive to the underlying parameters. 
We see immediately that the contribution of $\lambda_v$ to the variation of
$\mu$ is negligible; we have
\beq \label{muitounification}
 \Delta \ln \mu \simeq \left[ 13.0\lambda_X - (0.61\pm 0.12)\lambda_H + 0.35 \lambda_S \right] 
  \Delta \bar{\phi}.
\eeq

We can now for a range of unified models determine the relations between 
variations of $\mu$ and $\alpha$ and the low-energy coupling constants 
that give rise to differential accelerations. We ignore $\lambda_M$ since 
this coupling does not affect either $\alpha$ or $\mu$. 

\paragraph{Scenario 1: varying unified coupling}
We first consider the case when only $\lambda_X$ is nonvanishing, among the 
underlying cosmon couplings. Then we have 
\beq
 \Delta \ln \alpha \simeq 0.52 \lambda_X \Delta\bar{\phi},\qquad  
 \Delta \ln \mu \simeq 13 \lambda_X \Delta\bar{\phi},
\eeq
\beq
 \lae = 0,\qquad \lambda_i = 13 \lambda_X,\qquad \lambda_v = (-1.4\times 10^{-2}) \lambda_X
\eeq
thus $\Delta\ln\mu \simeq 25\Delta\ln \alpha$ (which would put the 
variation of $\alpha$ beyond the reach of current observations). 
The magnitude of $\lambda_X$ is at least of order $10^{-6}$, 
to produce a nonzero variation of $\mu$. Now 
the differential acceleration arises from the product $\lambda_i\lambda_v/2$ 
which is at least $10^{-13}$, leading to $\eta$ of magnitude just below 
$10^{-14}$; the contribution proportional to $\lambda_{em}$ is slightly 
smaller.

\paragraph{Scenario 2: varying Higgs v.e.v.\ without SUSY}
Next we consider if the variation were to arises solely from a change 
in the Higgs v.e.v.\ relative to the unified scale, thus $\lambda_H$ is 
nonzero. Then we have
\beq
 \Delta \ln \alpha \simeq (2\times 10^{-3}) \lambda_H \Delta\bar{\phi},\qquad  
 \Delta \ln \mu \simeq -0.6 \lambda_H \Delta\bar{\phi},
\eeq
\beq
 \lae = \lambda_H,\qquad \lambda_i \simeq 0.4 \lambda_H,\qquad 
 \lambda_v = 2\times 10^{-3}\lambda_H
\eeq
In this case the fractional variation of $\alpha$ should be about 
300 times smaller than that of $\mu$! Now to reproduce a nonzero 
variation of $\mu$ we require $\lambda_H$ to be a few times $10^{-5}$ 
and $\lambda_i\lambda_v/2$ is again at the level of $10^{-13}$, with 
the electron contribution $\lambda_i\lae /(2\cdot 1837)$ only slightly 
smaller. Again $\eta$ is bounded below by $10^{-14}$, given a nonzero 
variation of $\mu$. 

\paragraph{Scenario 3: variation of SUSY-breaking hidden sector and Higgs v.e.v.}
Next we consider what happens if supersymmetry-breaking arises from
a hidden sector with a mass scale that is generated by gauge dynamics, 
and assume that electroweak symmetry breaking is triggered by soft 
SUSY-breaking masses. We will first take the hidden sector gauge coupling
to vary independently of the visible sector gauge couplings, thus
\beq 
 \lambda_X = 0,\qquad \lambda_S \simeq \lambda_H.
\eeq
Then we find
\beq
 \Delta\ln\mu = (-0.3 \pm 0.1) \lambda_S \Delta\bar{\phi},\qquad  
 \Delta\ln\alpha = (1.3 \times 10^{-2}) \lambda_S \Delta\bar{\phi},
\eeq
\beq
 \lae \simeq \lambda_S, \qquad 
 \lambda_i \simeq 0.74 \lambda_S, \qquad 
 \lambda_v = 1.7 \times 10^{-3} \lambda_S,
\eeq
with $\lambda_S$ being at least $10^{-4}$ to produce a variation of 
$\mu$ at the observed level. Here the fractional variation of 
$\alpha$ should be about 20 times smaller than that of $\mu$ and of 
{\em opposite}\/ sign. Then the product $\lambda_i\lambda_v/2$
is at least a few times $10^{-12}$ and $\eta$ must reach the level
of $10^{-13}$ level, pushing current observational limits. The 
$\lae$ and $\lambda_{em}$ contributions are slightly smaller.

\paragraph{Scenario 4: varying unified coupling plus SUSY-breaking}
Finally we consider what may happen if the hidden sector gauge 
coupling varies with the unified coupling at the fundamental scale. 
As first pointed out in \cite{Langacker} this leads to the relations
\beq
 \lambda_S \simeq \lambda_H \simeq 34 \lambda_X.
\eeq
This results \cite{me03} in $\Delta \ln\mu$ being of the same 
magnitude as $\Delta\ln\alpha$. More concretely
\beq
 \Delta\ln\mu = (4\pm 4)\lambda_X \Delta\bar{\phi},\qquad 
 \Delta\ln\alpha \simeq 1.0 \lambda_X \Delta\bar{\phi}
\eeq
thus $\lambda_X$ is just below $10^{-5}$, or larger, to match 
the claimed nonzero variations. We then find 
\beq
 \lae \simeq 34 \lambda_X,\qquad 
 \lambda_i \simeq 38 \lambda_X, \qquad 
 \lambda_v \simeq (4\times 10^{-2}) \lambda_X,
\eeq
thus the products of couplings appearing in $\eta$ are a few times 
$10^{-11}$, leading to differential accelerations of order $10^{-12}$, 
already outside the experimental range! 

This comes about because the variations of $m_p/M_P$ and $m_e/M_P$ are 
large but cancel against one another in $\mu$; however the differential 
accelerations due to the nucleon mass splitting, the electron mass and
the nuclear electromagnetic energy do not in general cancel against
each other. The 
contributions from $\lambda_e$ and $\lambda_{em}$ are slightly smaller
than those from $\lambda_v$, and may have opposite sign, but even if 
they did cancel in one E{\" o}tv{\" o}s-type experiment this could not 
hold in other setups where source and test bodies had different values 
of $Z$ and $A$.
This scenario could only survive if the variations of $\alpha$ and $\mu$ 
were below currently claimed nonzero values. 

One can also consider a non-SUSY GUT scenario where the unification mass scale
and coupling are $M_X\simeq 10^{15}\,$GeV and $\alpha_X^{-1}\simeq 40$, where
the electroweak scale would arise from some technicolor-like sector, giving 
$\lambda_H= (31\pm 2)\lambda_X$ if the technicolor group is unified with the
Standard Model. The results for this type of model are in fact very close to 
Scenario 3. There is a cancellation in the (fractional) variation of $\mu$ 
such that is comparable to the variation of $\alpha$, however the scalar 
couplings are relatively large and $\eta$ is of order $10^{-12}$.

Models where there is a partial cancellation between variations of $m_e/M_X$ 
and $m_p/M_X$ have been put forward in \cite{me03} and the last reference 
of \cite{Calmet}, as a way of reconciling the claimed nonzero variations 
of $\alpha$ and $\mu$ which are incompatible in many simple unified theories.
However, these estimates of WEP violation show that such models are likely 
to have problems to respect current experimental bounds.

\paragraph{Simplified expression for WEP {\em vs.} varying mu} 
We have seen that in all these scenarios based on unification the source 
factor (first bracket) in Eq.~(\ref{etaitolambda}) is strongly dominated by the 
isospin-conserving scalar coupling $\lambda_i$, while the second factor 
depends on the electron coupling, the isospin-violating coupling $\lambda_v$
and the electromagnetic coupling $\lambda_{em}$. The isospin-violating 
contribution to $\Delta\ln\mu$ in Eq.~(\ref{muitolambda}) is negligible. 
Therefore the relation between the variation of $\mu$ and the differential 
acceleration $\eta$ can be simplified to 
\beq
 \frac{\eta}{(\Delta\ln\mu)^2} \gtrsim \frac{\lambda_i}{2(\lambda_i-\lae)^2} 
 \left( \left[ \frac{\lae}{1837} -\lambda_v \right] \Delta_{12} f_p
 + 7\times 10^{-4} \lambda_{em} 
 \Delta_{12} \frac{\langle Z^2 A^{-1/3} \rangle}{\langle A\rangle}
 \right)
\eeq
due to the fact that all scalar couplings arise from one underlying 
parameter whose variation, for any given $\Delta\ln\mu$, is proportional 
to $(\lambda_i-\lae)^{-1}$.

\section{Evolution of the cosmon or quintessence field}

Since the estimates of the previous section are interestingly close to the
current experimental limit of $\eta$, and we made only rough estimates using
the assumption that $\Delta\bar{\phi}$ was less than unity, we will refine 
them by examining in more detail the cosmological evolution of $\phi$
and applying constraints on its kinetic energy density.
In previous work \cite{Wetterich02} a physical model was 
adopted with an exponential potential and a general (non-canonical) kinetic 
term allowing the past evolution of the field to be found in terms of a few 
parameters. However in general the evolution and variation of ``constants'' 
may depend on undetermined {\em functions}\/ of a scalar field, either in 
the scalar action (kinetic terms plus potential) or its couplings to matter.

Another question is whether the evolution is homogeneous, that is, whether 
the scalar value depends only on cosmological time, or is position- or 
environment-dependent. This question was investigated for gravitationally 
collapsing regions of spacetime in \cite{BarrowMota} and more recently for 
virialized structures in \cite{BarrowShaw}, where the result was found that 
the rate of change of $\phi$ is the same within virialized regions as in the 
background and the value of the field depends only weakly on position, a 
result also obtained in \cite{BarrowSpatial,Avelino:2005pw} (confirming the 
analysis of \cite{Wetterich02}). However, note that if a scalar 
field couples strongly to matter such that its expectation value and mass are 
determined by the local density, both the cosmological evolution and
the possible scalar-mediated forces behave quite differently \cite{chameleon,
newhorizons}. It has been claimed that such a ``chameleon'' field cannot give 
rise to a fractional variation of $\alpha$ at the $10^{-5}$ level, but that
bound may be due to particular choices made in setting up the model of 
\cite{chameleon}, rather than a general result. We will consider here
only the weak coupling regime where the mass of the scalar field is 
determined primarily by its potential $V(\phi)$.

\subsection{Relating WEP violation to atomic clock measurements}
The $\phi$ couplings to matter can be related most directly 
to the {\em present}\/ rate of change of dimensionless couplings, which we now derive.
For a canonically normalized scalar field we have kinetic energy $T=(1/2)
\dot{\phi}^2$ and a potential $V(\phi)$, resulting in the energy 
density $\rho_\phi = T+V$ and equation of state $w_\phi = (T-V)/(T+V)$. 
There may also be contributions to the scalar equation of motion from 
interactions with matter, dark matter {\em etc.} and as shown for example 
in \cite{chameleon} these may alter the effective equation of state to 
be less negative. However we do not expect such contributions
to greatly affect our bound, which is derived from the maximum kinetic 
energy of $\phi$ consistent with observation and does not depend 
strongly on $w_\phi$.

The energy density fraction $\Omega_\phi$ is defined as 
$\rho_\phi/\rho_c \equiv \rho_\phi/3H^2\bar{M}_P^2$. Hence we obtain 
\beq \label{phidotonH}
 \frac{\dot{\bar{\phi}}}{H} = \sqrt{3\Omega_\phi(1+w_\phi)}.
\eeq
The usefulness of this equation is that the Hubble constant is now well 
measured whilst the quantities inside the square root on the RHS are 
bounded above. Given current cosmological observations, there is an allowed 
region in the $\Omega_\phi$-$w_\phi$ plane inside which Eq.~(\ref{phidotonH}) 
attains a maximum value, which we can estimate roughly as $\sqrt{3\times 
0.75\times 0.2} \simeq 0.7$. 
Using $H_0\simeq 7\times 10^{-11} y^{-1}$ we find
\beq
\dot{\bar{\phi}} \leq \dot{\bar{\phi}}_{\rm max} \simeq 5\times 10^{-11} y^{-1}.
\eeq
For a given unified model where all varying quantities are determined by a 
single parameter $\lambda$ we can write 
\beq
 \frac{\dot{\alpha}}{\alpha} \leq c_1 \lambda \dot{\bar{\phi}}_{\rm max},\qquad
  \frac{\dot{\mu}}{\mu} \leq c_2 \lambda \dot{\bar{\phi}}_{\rm max},
\eeq
where the numerical constants $c_1$ and $c_2$ arise from relations 
detailed in the previous section. Similarly the results for $\eta$ can
be summarized in the form 
\beq
\eta = K \frac{\Delta_{12} f_p}{2}\lambda^2
\eeq
for some constant $K$, where for simplicity we omit the contribution 
of nuclear electromagnetic energy which was subdominant in all the 
scenarios considered.~\footnote{With an experiment using test bodies of 
very large $Z$, this contribution could become competitive or dominant.}
Thus we find
\beq
 \eta \geq \frac{\Delta_{12}f_p}{2 \dot{\bar{\phi}}_{\rm max}^2} 
 \frac{K}{c_2^2} \left(\frac{\dot{\mu}}{\mu}\right)^2
 \simeq \frac{K}{c_2^2} \left(\frac{\dot{\mu}/\mu}{3.7\times 10^{-10}y^{-1}} \right)^2
\eeq
taking $\Delta_{12}f_p$ to be $0.036$ as before. An exactly analogous 
inequality follows by replacing $\mu$ with $\alpha$ and $c_2$ with $c_1$. 

Now if we suppose that $\dot{\alpha}$ or $\dot{\mu}$ are nonzero at present 
(either due to direct measurement or to a particular model of cosmological 
evolution) then in any given scenario of unification we find the values 
of $K/c_1^2$ and $K/c_2^2$ and obtain the lower bound on $\eta$.

Conversely, given bounds on $\eta$, for any given unification scenario we 
can obtain bounds on $\dot{\alpha}/\alpha$ and $\dot{\mu}/\mu$ which must 
be satisfied for a cosmological variation arising from a scalar field 
effective action. These may be directly compared with atomic clock 
measurements. Note that a nonzero result for $\eta$, while extremely
interesting, would not have direct implications for ``varying constants''
within this framework since the rate of change of $\phi$ is not bounded 
below.

As an example we evaluate such bounds within Scenario 1, where the
relevant scalar coupling is $\lambda_X$ and the numerical constants are 
$K/c_1^2 \simeq 0.67$ and $K/c_2^2 \simeq 0.0011$. 
Thus for a hypothetical variation $\dot{\alpha}/\alpha \sim 10^{-16}y^{-1}$, 
we find that $\eta$ should be at least $5\times 10^{-14}$.

For Scenario 4, which is barely ruled out from the point of view of
producing the claimed nonzero variations of $\alpha$ and $\mu$, we find
the larger values $K/c_1^2 \simeq 1.5$ and $K/c_2^2 \simeq 0.1$ reflecting 
the fact that such scenarios produce small $\mu$ variations through 
partial cancellation. 

\subsection{Bound on variation of $\bar{\phi}$ at earlier times}
Equation (\ref{phidotonH}) may also be extended to earlier times, given 
bounds on the energy content of the universe derived from cosmological 
measurements. Changing variables from time to redshift we find for the 
change in the scalar field value back to some redshift $z$
\beq
 |\Delta \bar{\phi}| \leq \ln(1+z)\ 
 \mbox{max}\left(\sqrt{3\Omega_\phi(1+w_\phi)} \right)
\eeq
where the maximum is to be evaluated over the range from $z$ to the present.
A very conservative bound is a scalar field behaving like cold dark matter, 
for which $w=0$ and $\Omega\leq 0.35$, thus the square root is $1$ or less. 
This leads to a limit such as $|\Delta \bar{\phi}| \leq 0.12$ at the Oklo 
epoch $z=0.13$ (see the last reference of \cite{Oklo}), or $|\Delta 
\bar{\phi}| \leq 1.4$ at redshift 3. This justifies the previous assumption 
that $|\Delta \bar{\phi}|$ was at most of order 1 at redshifts relevant for 
QSO absorption spectra. We emphasize that in almost all viable scalar 
theories such values of $\Delta \bar{\phi}$ will be a considerable 
overestimate, since it would require considerable fine tuning to saturate 
these upper bounds on $\dot{\bar{\phi}}$ at all times.

In theories where the nonlinear dependence of coupling strengths on the 
canonically normalized scalar $\bar{\phi}$ is known, such results can be 
used to check consistency of measurements of $\alpha$, $\mu$ and other
quantities at different epochs. In general as stated earlier the 
dependence is allowed to be an arbitrary function of $\bar{\phi}$ and 
only the derivative at the current epoch is probed by WEP and atomic clocks.

\section{Conclusions}
We have investigated 
bounds arising from 
violations of the Weak Equivalence Principle in scalar field theories in 
which the proton-electron mass ratio $\mu$ may vary. We related such bounds 
to unified theories, astrophysical observations of quasar absorption spectra 
and atomic clock measurements of the constancy of physical parameters. 
Model-independent bounds on the rate of change of the scalar field and on 
its total variation at any given epoch are obtained from the effect of its 
kinetic energy on the expansion of the Universe. 

For a given nonzero variation of $\mu$, the scalar couplings to the proton 
or electron mass are bounded below, thus so is the resulting differential 
acceleration, up to a single parameter $K/c_2^2$. This parameter is a ratio 
of scalar coupling constants, which arises from how the nucleon and electron
masses are related in the underlying particle theory. Hence such underlying 
theories, in particular gauge unification at high energy, could be tested with 
variation of $\mu$.

Since the value of $\alpha$ may also vary in such underlying theories, we 
also analyzed its variation in relation to WEP violation, extending previous 
works. We found that in many scenarios for generating varying $\mu$ from unified
models, the differential acceleration parameter $\eta$ is near or above its
current experimental limit. In particular, in scenarios where due to 
cancellations the variation of $\mu$ is not much larger than that of $\alpha$,
the contributions to differential accelerations do not cancel and the 
violation of WEP is at or above the experimental limit. Thus it seems very 
difficult to reconcile currently claimed nonzero variations of $\mu$ and 
$\alpha$ within unified models.

Our results differ from those of \cite{BarrowElectron}, where the 
cosmological variation of $\mu$ was bounded to be many orders of magnitude 
below current levels of sensitivity. The main difference is that in 
\cite{BarrowElectron} the source of the scalar field variation was taken 
to be the coupling to the local electron density, whereas we allow 
the field to be driven by a potential energy and in principle by couplings 
to other constituents of the Universe. This illustrates the point that in 
models where the form of the scalar Lagrangian is known or assumed, more 
stringent bounds may be found. Our aim was however to find bounds which are
as model-independent as possible.

\subsection*{Acknowledgments}
Helpful discussions with C.~Wetterich, M.~Doran, M.~Murphy, D.~Mota and 
D.~Shaw are acknowledged. The author is supported by the {\em Impuls- and 
Vernetzungsfond der Helmholtz-Gesellschaft}.



\end{thebibliography}

\end{document}